\documentclass[conference]{IEEEtran}

\IEEEoverridecommandlockouts
\usepackage{cite}
\usepackage{amsmath,amssymb,amsfonts}
\usepackage{algorithmic}
\usepackage{graphicx}
\usepackage{acronym}
\usepackage{textcomp}
 \usepackage{balance}
\usepackage{wrapfig}
 \usepackage{booktabs}

\usepackage{caption}
\usepackage{subcaption}
\newcommand{\figref}[1]{fig. \ref{#1}}
\usepackage{xcolor}
\usepackage{tabularray}
\usepackage{multirow}

\usepackage{eurosym}

\def\BibTeX{{\rm B\kern-.05em{\sc i\kern-.025em b}\kern-.08em
    T\kern-.1667em\lower.7ex\hbox{E}\kern-.125emX}}
\usepackage[hyphens]{url}
\usepackage[hidelinks]{hyperref}
\hypersetup{breaklinks=true}
\begin{document}
\title{Aggregate Peak EV Charging Demand: The Influence of Segmented Network Tariffs
\thanks{The research was supported by the ROBUST (MOOI32014) and GO-e (MOOI32001) projects, which received funding from the MOOI subsidy programme by the Netherlands Ministry of Economic Affairs and Climate Policy and the Ministry of the Interior and Kingdom Relations, executed by the Netherlands Enterprise Agency.}}

\author{
\IEEEauthorblockN{Nanda Kishor Panda, Na Li, Simon H.~Tindemans}
\IEEEauthorblockA{Department of Electrical Sustainable Energy\\
Delft University of Technology\\
Delft, The Netherlands}
\{n.k.panda, n.li-2, s.h.tindemans\}@tudelft.nl
}

\maketitle

\begin{abstract}
Aggregate peak Electric Vehicle (EV) charging demand is a matter of growing concern for network operators as it severely limits the network's capacity, preventing its reliable operation. Various tariff schemes have been proposed to limit peak demand by incentivizing flexible asset users to shift their demand from peak periods. However, fewer studies quantify the effect of these tariff schemes on the aggregate level. In this paper, we compare the effect of a multi-level segmented network tariff with and without dynamic energy prices for individual EV users on the aggregate peak demand. Results based on real charging transactions from over 1200 public charging points in the Netherlands show that the segmented network tariff with flat energy prices results in more diverse load profiles with increasing aggregation, as compared to cost-optimized dispatch based on only dynamic day-ahead energy prices. When paired with dynamic energy prices, the segmented tariff still outperforms only dynamic energy price-based tariffs in reducing peaks. Results show that a balance between power thresholds and price per threshold is crucial in designing a suitable tariff, taking into account the needs of the power network. We also provide valuable insights to network operators by calculating the diversity factor for various peak demands per charging point.

\end{abstract}

\acrodef{EV}{Electric Vehicle}
\acrodef{CPO}{Charging Point Operator}
\acrodef{CS}{Charging Station}
\acrodef{CP}{Charging Point}
\acrodef{DSM}{Demand Side Management}
\acrodef{DSO}{Distribution System Operator}
\acrodef{OCPI}{Open Charge Point Interface}
\acrodef{TSO}{Transmission System Operator}

\begin{IEEEkeywords}
Aggregate, Distribution system, Electric vehicle, Flexibility, Segmented tariff
\end{IEEEkeywords}
\section{Introduction}

\IEEEPARstart{E}{lectric} Vehicles offer an essential shift in the transportation sector by offering a sustainable alternative to internal combustion-based engines. This transition aligns with the global effort to reduce the dependence on fossil fuels and thus contribute to the energy transition~\cite{yuan2021electrification, barman2023renewable}. The adoption of \acp{EV} has resulted in an extensive network of \ac{EV} chargers, primarily connected to the low voltage grid. With the addition of more \acp{EV}, their aggregate peak demand poses a threat to the operational reliability of the power networks through network congestion~\cite{das2020electric, venegas2021active}.

Current low voltage grids are already operating at their peak capacity, unable to accommodate additional loads without timely reinforcement~\cite{verbist2023impact}. The majority of grid congestion primarily occurs when periods of high demand from non-\ac{EV} loads coincide with peak charging demand for \acp{EV}. For instance, residential \acp{CS} experience a surge in charging demand during evening hours, aligning with the peak demand for houses where individuals return from work and charge their \acp{EV}. \par

The charging flexibility of \acp{EV} offers an opportunity to shift or distribute their charging demand over the connection period, which is typically much longer than the minimum time required to charge them~\cite{panda2024quantifying}. However, the lack of standardized incentive schemes limits the potential of aggregate \ac{EV} charging. For example, charging based solely on energy prices can result in new peaks during periods of low prices instead of eliminating them~\cite{van2018flexibility}. Additionally, there is limited research on the effect of network tariffs on the aggregate peak \ac{EV} charging demand.

As a solution to this problem, researchers have proposed leveraging \ac{EV} flexibility to reduce peaks during hours of high demand through peak shifting, valley filling, or peak shaving~\cite{eid2016time}. These demand response methods can either be applied centrally through a central agent or individually at the device level. Centralized control and scheduling of \acp{EV} can result in optimal schedules but requires sophisticated mathematical models and powerful optimizers. On the other hand, individual \ac{EV} charging can be optimized individually based on a signal, which is often based on prices. The prices can comprise varying energy prices and constant network tariffs~\cite{daneshzand2023ev}, or a combination of varying energy prices and network tariffs~\cite{verbist2023impact, fitzgerald2017evgo}. The authors in~\cite{yong2023electric} provide a comprehensive overview of dynamic charging tariffs for charging stations, such as time-of-use pricing, critical peak pricing, and real-time pricing. All these studies show the importance of using additional network tariffs apart from energy prices to be able to manage \ac{EV} charging peaks and solve congestion problems. \par 

Scheduling \acp{EV} based on only energy prices can result in lower charging costs but may exacerbate coincidence of peak demand at the aggregate scale. One of the ways to limit the aggregate peak charging demand is by restricting individual charging power levels. Various researchers have implemented this through the addition of a new tariff component based on power, which can replace the conventional flat/fixed network tariffs.\par

The work in~\cite{tuunanen2016power} explores the concept of power-based distribution tariffs for distribution system operators by charging customers based on peak power usage. It highlights the benefits of using power-based tariffs to incentivize customers to reduce peak demand. \cite{rautiainen2017reforming} discusses the transition towards power-based tariffs. It proposes alternative, more cost-reflective tariff structures like the power tariff, threshold power tariff, power limit tariff (also known as power band tariff), and step tariff. The power tariff consists of three cost components: basic charge (in €/month), energy charge (in €/kWh), and power charge (in €/kW) based on peak power (i.e. the highest or the there highest hourly power of the month). The threshold power tariff has a similar cost component but applies the power charge only when consumption exceeds a predefined threshold. This tariff structure and its implications for reducing peak demand and ensuring \acp{DSO} cost recovery have been further examined in studies in~\cite{lummi2016cost, lummi2016variations, lummi2017analysis}. The power limit tariff simplifies to a single power charge, where consumers pre-select a maximum power level and are penalized for exceeding this limit, a concept aligning with capacity subscription tariffs discussed in the work of~\cite{van2019variable, van2021designing, hennig2020capacity}. Conversely, the step tariff uses a basic charge (in €/month) and a consumption charge (in €/kWh). If the average power remains within a certain predefined limit, then the charge is low; otherwise, the charge is very high. Another type of power-based tariff - segmented tariff, is proposed in~\cite{li2020segmented}; it uses a consumption charge (in €/kWh), where it assigns a tariff to each power threshold, the higher the threshold is, the higher the tariff is. The results show that this method can efficiently flatten the aggregate load profile in the case of residential users with energy storage. Further~\cite{li2023residential} shows how a multi-level segmented network tariff can do a better job in flattening peaks along with cost recovery for \acp{DSO}.\par

\ac{EV} charging is similar to conventional load consumption but with more flexibility. Also, due to the small size of batteries and distributed usage, the full potential of \ac{EV} flexibility can only be realized at an aggregate level. Hence, it is important to assess the impact of different tariff structures on the aggregate peak charging demand of \acp{EV}. However, limited studies exist to assess the effect of tariffs based on power levels when used for \ac{EV} charging and how its effect scales with aggregation.\par

In this study, we study aggregate \ac{EV} charging demand using real \ac{EV} charging session data. We compare unoptimized charging with cost-minimized charging schemes based on day-ahead hourly prices, and the effects on aggregate peak demand. For each of these base scenarios, we investigate the effectiveness of segmented network tariffs on mitigating aggregate peak \ac{EV} charging demand. 
Furthermore, we quantify the aggregate peak demand of \ac{EV} fleets using relevant metrics such as the peak-charging power per connection and the diversity factor, for different aggregation levels.\par

The remainder of this paper is organized as follows: Section \ref{modeling} describes the models and formulates the decision problems of EV owners mathematically. Results are analyzed in Section \ref{Results analysis}. Finally, Section \ref{Conclusions} concludes the paper by providing future work recommendations.
\begin{table*}
\centering
\captionof{table}{Overview of different analyzed tariffs}\label{tab:tariff-alias}
\begin{tabular}{cclc} 
\toprule
\textbf{Tariff type}                                                                  & \textbf{Alias}      & \multicolumn{1}{c}{\textbf{Power levels~(kW)}} & \textbf{Prices~(€/kWh)}                                             \\ 
\bottomrule
\textbf{Unoptimized (~\ref{unoptimized})}                                                                & Unopt               & \multicolumn{1}{c}{-}                          & -                                                                   \\ 
\midrule
\textbf{Dynamic energy~pricing (\ref{cost-opt}-\ref{dynamic})}                                                   & DE                  & \multicolumn{1}{c}{-}                          & -                                                                   \\ 
\midrule
\multirow{2}{*}{\textbf{Segmented network tariff with flat energy price (\ref{cost-opt}-\ref{flat+seg})}}             & FE-$p^+$          & $\bar{p}_{0,1,2}=\{4,8,11\}$                     & \multirow{2}{*}{-}                                                  \\ 
\cmidrule{2-3}
                  & FE-$p^-$          & $\bar{p}_{0,1,2}=\{2,4,17\}$                     &                                                                     \\ 
\cmidrule{2-3}
\multirow{4}{*}{\textbf{\textbf{Segmented network tariff with dynamic energy price (\ref{cost-opt}-\ref{dynamic+seg})}}} & DE-$p^+\lambda^-$ & \multirow{2}{*}{$\bar{p}_{0,1,2}=\{4,8,11\}$}  & \multicolumn{1}{l}{$\bar{\lambda}_{0,1,2} = \{0, 0.055, 0.900\}$~}  \\ 
\cmidrule{2-2}
                                      & DE-$p^+\lambda^+$  &                                                & \multicolumn{1}{l}{$\bar{\lambda}_{0,1,2} = \{0, 0.158, 0.900\}$}   \\ 
\cmidrule{2-3}
                                      & DE-$p^-\lambda^-$  & \multirow{2}{*}{$\bar{p}_{0,1,2}=\{2,4,17\}$}  & \multicolumn{1}{l}{$\bar{\lambda}_{0,1,2} = \{0, 0.055, 0.900\}$}   \\ 
\cmidrule{2-2}
                                                                                      & DE-$p^-\lambda^+$ &                                                & \multicolumn{1}{l}{$\bar{\lambda}_{0,1,2} = \{0, 0.158, 0.900\}$}   \\
\bottomrule
\end{tabular}
\end{table*}
\section{Methods and Models} \label{modeling}
\subsection{Overview of charging topology}

This paper considers the impact of pricing incentives on \ac{EV} charging. For the purpose of this paper, we restrict the analysis to \ac{EV} chargers that have their own grid connection, which does not need to be shared with other loads or injections. In this context, an \ac{EV} user's total charging costs consist of energy and network costs. Energy costs refer to the actual energy charged, while the network costs cover both \ac{TSO} and \ac{DSO} costs that these operators incur to maintain and operate the network infrastructures~\cite{ACER2023tariff}. 

Presently, in the Netherlands, \ac{EV} users who have a private charging point within their residence pay for the energy they consume at the same rates as their regular household electricity consumption. These energy prices can be dynamic (based on day-ahead prices), varying based on time-of-use, or, in some cases, can be fixed~\cite{ANWB2030EV}. Along with the energy costs, the users also pay fixed network costs based on their connection capacity. The users of public \ac{EV} chargers only pay for the energy used based on a fixed price per kWh. Currently, this price is determined by the municipality after consultation with the \acp{CPO}. \acp{CPO} cover energy expenses based on day-ahead market rates and fixed network costs corresponding to their connection capacities. Their profit margin is the difference between the expenses covered by \acp{CPO} and the revenue from \ac{EV} users~\cite{BRINKEL2023100297}.

In this paper, public \acp{CS} operated by a single \ac{CPO} are analyzed. These  \acp{CS} usually consist of two \acp{CP}, which are the physical outlets where an \ac{EV} can be connected. For the analyzed data, each \acp{CS} can provide up to 23 kW of power distributed among two \acp{CP}, each with a maximum power output of 23 kW.  In the following, we analyze the results per \ac{CP} in order to normalize their usage pattern. The data is analyzed for 2022 with approximately 300,000 charging transactions spread over 650 \acp{CS}. A single charging transaction is characterized by maximum charging power ($\bar{p}$) and energy ($\bar{e}$) that have to be charged between its arrival ($t^a$) and departure ($t^d$) time. Different dispatch strategies are simulated for each of the charging transactions based on historical data and then aggregated to quantify the desired matrices.\par

\subsection{Electricity pricing}

In this paper, we consider two types of electricity pricing: one is flat energy pricing, which is constant over the whole year, and the other is dynamic energy pricing, which varies with each hour. In this paper, we consider the hourly day-ahead energy prices for 2022 in the Netherlands, provided by~\cite{ENTSO-E}. This omits taxes and additional fees imposed by the energy supplier. Note that price data is taken from the same year as the charging transactions, ensuring the correct dependence on the charging sessions from the same year. 

\subsection{Network tariffs}
\begin{figure*}
    \centering
    \begin{minipage}[t]{0.35\textwidth}
         \centering
         \includegraphics[width=0.63\linewidth]{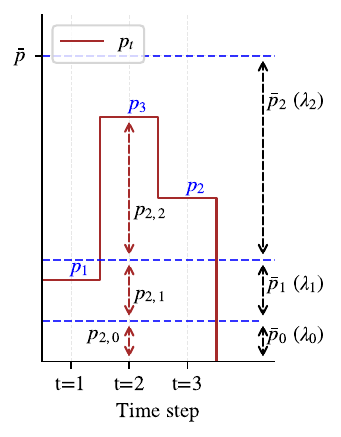}
         \caption{Illustration of three-level segmented tariff. The three thresholds of the segmented tariff are denoted by $\bar{p}_{0,1,2}$ along with their respective prices ($\lambda_{0,1,2}$).}
         \label{fig:segmented_tariff}
     \end{minipage}
     \hfill
     \begin{minipage}[t]{0.6\textwidth}
     \centering
          \includegraphics[width=\linewidth]{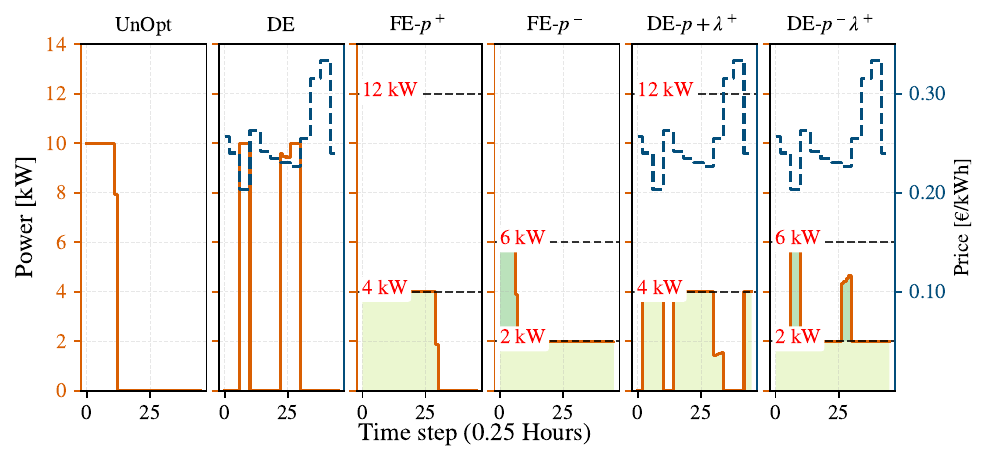}
         \caption{Representative charging profiles illustrating different dispatch strategies for a single \ac{EV}. The low-price ($\lambda^-$) variants of segmented tariffs are not shown because the results are identical to the high-price ($\lambda^+$) variants for this session. }
         \label{fig: resutls-1}
     \end{minipage}
\end{figure*}

Two types of network tariffs are modelled. The first is a fixed connection fee, regardless of consumption behaviour (up to the contractual limit). The second type is a segmented network tariff that introduces an additional cost structure tied to power consumption~\cite{li2020segmented}: users pay an additional volumetric fee (€/kWh) for their power consumption above a threshold. Multiple thresholds can be used to create an escalating fee structure. This encourages users to flatten their load profiles, thus preventing possible congestion in the distribution network~\cite{li2023residential}. 
The segmented network tariff with three power levels is illustrated in~\figref{fig:segmented_tariff} for three time steps. In this illustration, the power-related network costs for time step 2 amount to $\sum_{s={0,1,2}} \lambda_s p_{2,s}$, where $p_{2,s}$ represents the power utilized within the $\bar{p}_s$ segment during time step 2, and $\lambda_s$ signifies the network price (€/kWh) assigned to that specific power segment. In this model, $\lambda_0$ represents a base volumetric fee that applies even to low power consumption. It may be set to zero. 

A network tariff, with or without segmentation, may also contain a consumption-independent (fixed) component. The magnitude of this component is important for cost recovery of the network owner/operator~\cite{li2023residential}. However, it is not considered for the analysis in this paper because it does not impact the charging schedule.

\subsection{Dispatch strategies} \label{sec:dispatch_strat}

Three different cost-optimizing dispatch strategies are modelled to assess their impact on the aggregate peak of \ac{EV} charging power. Their performance is compared to the unoptimized charging strategy, also known as `dumb charging'. All strategies are explained below.\par

\subsubsection{Unoptimized charging} \label{unoptimized}
The unoptimized charging strategy is the default strategy, which charges the \ac{EV} with the maximum power as soon as it is connected until the \ac{EV}'s battery is charged to the desired level. Unoptimized charging is modeled as a linear program using~\eqref{eq:dumb_obj}-\eqref{eq:base_constraint_charging_power}:
    \begin{align}
        \max_{p} \sum_{t\in \mathcal{T}} e_{t} \label{eq:dumb_obj}
    \end{align}
    subject to:
    \begin{subequations}
    \label{eq:allconstraints}
    \begin{align}
    e_{t} &= 0,  &&& t\leq t^a \label{eq:constraint-2}\\  
    e_{t} &= e_{t-1}+p_{t-1}\Delta t, &&& t^a <t<t^d \\
    e_{t} &= \overline{e}, &&&  t\geq t^d \label{eq:base_constraint_charging_energy}\\
    p_{t} &= 0, &&& t<t^a \wedge  t\geq t^d\\
    0 &\leq  p_{t} \leq \overline{p},  &&& t^a\leq t < t^d    \label{eq:base_constraint_charging_power} 
    \end{align}
    \end{subequations}

\subsubsection{Cost optimized charging}\label{cost-opt}
Minimizing the cost of charging is a natural objective for \ac{EV} users exposed to price incentives. In this paper, we present three dispatch strategies based on minimizing charging costs, as shown below:
\begin{enumerate}
    \item \emph{Charging with dynamic energy prices:} \label{dynamic} Dynamic energy prices ($\lambda_t^e$), such as day-ahead prices, vary with time and can be used to incentivize charging during periods when energy is produced at low (marginal) cost. Here, we assume the network costs are fixed and are not considered in the optimization model.
\begin{align}
    \min_{p}\:  \sum_{t\in \mathcal{T}}  \lambda_t^e p_{t} \Delta   t \quad\quad\text{subject to: }  \eqref{eq:allconstraints}    \label{eq:obj_dc}
\end{align}
This yields a unique solution when prices are not identical for different time steps. 

\item \emph{Segmented network tariff with flat energy price:} \label{flat+seg}
When a connection with an \ac{EV} charger is exposed to segmented network tariffs and flat energy prices, the optimal charging pattern is computed by solving the problem:
\begin{equation}
    \min_{p}\:  \sum_{t\in \mathcal{T}} \sum_{s\in \mathcal{S}}\lambda_s p_{t,s}-\epsilon \sum_{t\in \mathcal{T}} e_{t}
    \label{eq:obj_dn_fc}
\end{equation}
subject to: \eqref{eq:allconstraints} and
\begin{subequations}    
\label{eq:additionalconstraints}
\begin{align}
p_{t} &= \sum_{s\in \mathcal{S}} p_{t,s} &&& \forall t\in \mathcal{T} \label{eq:con_power_seg}\\
0 &\leq p_{t,s} \leq \bar{p}_{s}  &&&\forall s=\{0, \cdots,|\mathcal{S}|-1\}, t\in \mathcal{T}\\ 
\lambda_s &\leq \lambda_{s+1}  &&& \forall s \in \mathcal{S} \label{eq:con_segment}
\end{align}
\end{subequations}
As flat energy prices have no effect on the optimal charging schedule, they are excluded from the optimization model. An additional term $\epsilon \sum_{t\in \mathcal{T}} e_{t}$ is included in the objective function \eqref{eq:obj_dn_fc}, with a small coefficient $\epsilon$. This secondary objective ensures that - among equally costly solutions - the one that completes charging first is preferred. For sufficiently small values of $\epsilon$, this also guarantees the uniqueness of the solutions.

\item \emph{Segmented network tariff with dynamic energy prices:}\label{dynamic+seg}
Finally, flexible \ac{EV} users may be exposed to segmented network tariffs and dynamic energy prices at the same time. The resulting model, combining both objectives, is shown below:
\begin{align}
    \min_{p}\:  \sum_{t\in \mathcal{T}} \sum_{s\in \mathcal{S}}\lambda_s p_{t,s} + \sum_{t\in \mathcal{T}} \lambda_t^e p_t \Delta t
\end{align}
subject to: \eqref{eq:allconstraints} \&
\eqref{eq:additionalconstraints}. The $\epsilon$-weighted additional objective is not included here, due to the greater variation of price levels within a single charging session.

\end{enumerate}

\section{Results} \label{Results analysis}

\subsection{Individual and aggregate charging profiles} 
The interaction between grid tariff and electricity price incentives is sensitive to the choice of the capacity thresholds $\overline{p}_i$ and the relation between the surcharges $\lambda_i$ and the electricity price signal. For this reason, two scenarios are implemented for each of these, resulting in the charging scenarios that are summarized in Table~\ref{tab:tariff-alias}. The price level $\lambda_1$ for the middle threshold ($\bar{p}_1$)  band was set to the $5\textsuperscript{th}$ and $25\textsuperscript{th}$ quantile of the electricity prices in the hourly price series for the year 2022. The base level $\lambda_0$ was set to zero and $\lambda_2$ to 0.90~€/kWh. The latter exceeds the maximum price seen in the data and thus serves as a strong deterrent to high-power charging without strictly curtailing users.

The charging profiles for a representative \ac{EV} charging session in combination with incentives scenarios are shown in~\figref{fig: resutls-1}. Where applicable, power levels for the segmented network tariff and dynamic energy prices are indicated. The presented \ac{EV} transaction has a maximum charging power of 11 kW and a charging demand of 60 kWh for a connection duration of 12.25 hours. The results illustrate that (1) implementing a dynamic energy pricing strategy effectively shifts peak demand from high-price to low-price hours, (2) a segmented tariff combined with a flat energy price significantly reduces peak demand if there is sufficient time to charge, and (3) when a segmented tariff is paired with dynamic energy pricing, the outcomes depend on the relative levels of the tariff and energy prices.

\begin{figure}
    \centering
    \includegraphics[width=\linewidth]{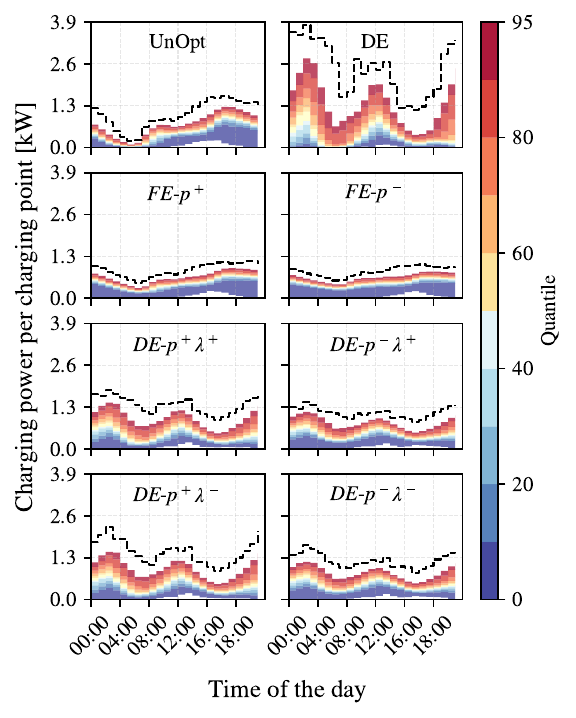}
    \caption{Quantile distribution of charging demand for different dispatch strategies across all days in the year 2022 for all sets of \acp{CP}. Colours indicate the quantiles of charging power for each hour relative to all days in the year. The dashed lines indicate the maximum observed charging power for each hour.}
    \label{fig:agg_quantile}
\end{figure}

\begin{figure*}[!ht]
\centering
\includegraphics[width=13cm]{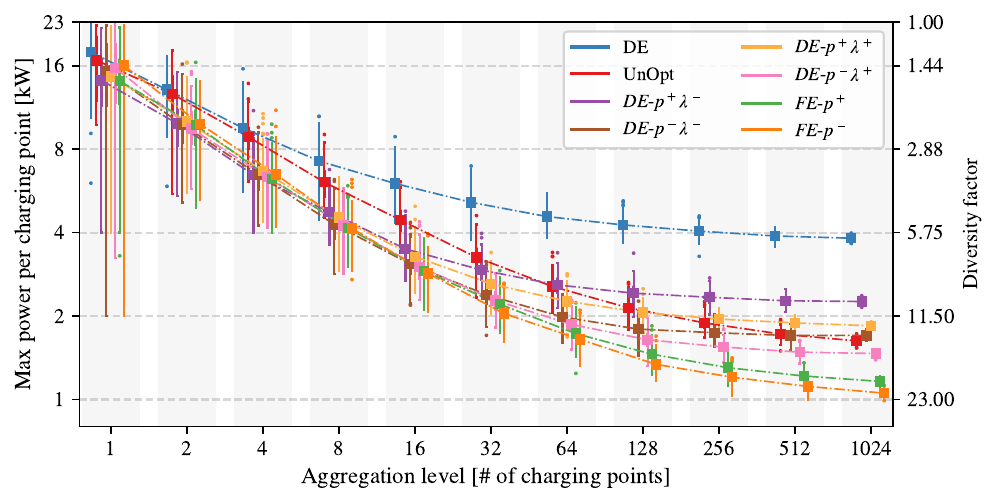}
\caption{Distribution of annual maximum power per charging point under different aggregation levels for different dispatch strategies. The diversity factor corresponding to each power level is also indicated on the right axis.}\label{fig:diversity_factor}
\end{figure*}

The quantile distribution of charging power per \ac{CP} over the year for different dispatch strategies are presented in~\figref{fig:agg_quantile}. Optimizing the individual charging schedule based on dynamic energy (DE) prices (here, day-ahead prices) results in larger peaks, surpassing the unoptimized case. This is due to the scheduling of charging at the same moment when prices are low. The maximum average peak for the case of dynamic energy tariff can increase by as much as 16\% when compared with the unoptimized case. Hence, using cost-minimized charging based only on energy costs can result in higher aggregate power peaks. \par
Segmented network tariffs with flat energy tariffs (FE-$p^+$ \& FE-$p^-$) result in a flatter charging profile at the aggregate level due to additional costs based on the power level of individual \acp{EV}. As segmented tariffs are sensitive to different power thresholds, we analyzed two power levels, one with a higher (FE-$p^+$) threshold than the other (FE-$p^-$). As expected, the aggregate charging profile is flatter for the low threshold case.

We also examined the cases of segmented network tariffs with dynamic energy prices (DE-$p^{\{+,-\}}\lambda^{\{+,-\}}$). Although, in this case, there are considerable peaks in the aggregate charging profile compared to the case with flat energy prices, profiles are still much flatter than in the dynamic energy or unoptimized case. With the addition of a dynamic energy price component, the performance of segmented network tariffs is now sensitive to the relative value of the energy prices and prices of power thresholds. This dependence is illustrated by simulating two prices ($\lambda^+$ and $\lambda^-$), based on the 5\textsuperscript{th} and 25\textsuperscript{th}  quantile values of the yearly day-ahead price, respectively. In cases where the energy price difference within a charging session exceeds the cost of charging above a tariff threshold, the \ac{EV} may opt to charge with a higher power, resulting in differences between the $\lambda^+$ and $\lambda^-$ scenarios. 

Comparing the cases of segmented network tariffs with and without flat energy prices, it is clear that the power levels are the primary determinant affecting aggregate power consumption, and the price level for the middle capacity band has a lesser impact. It should be noted that in all cases, the price level for the upper capacity band exceeds the dynamic energy prices.

\subsection{Quantifying aggregate peak \ac{EV} charging demand}



The peak electricity demand of \acp{CS} at various aggregation levels is of particular concern for distribution grid owners, because it is a key driver for asset investments. To assess peak demand, we randomly selected $N$ \acp{CP} (without replacement) and generated a load profile from all sessions linked to those \acp{CP}. This process was done for aggregation levels $N=1,\ldots,1024$, and for each incentive combination. Fig.~\ref{fig:diversity_factor} shows the peak annual power consumption (normalised by the number of \acp{CP}) across 15-min intervals, including the statistical variation of this value across 100 random selections of \acp{CP} from the dataset. 

The \emph{diversity factor} is a useful measure of the diversity of charging behaviour across the population~\cite{chukwu2014impact, sun2018probabilistic}. It is defined as 
\begin{equation}
    d = \frac{\max_t p^{\mathrm{single}}_t}{\max_t p^{\mathrm{agg}}_t} = \frac{23~\mathrm{kW}}{\max_t p^{\mathrm{agg}}_t} ,
\end{equation}
where $p^{\mathrm{agg}}_t$ is the simulated aggregate power signal and $(\max_t p^{\mathrm{single}}_t )=23~\mathrm{kW}$, i.e. the network capacity required for a single \ac{CP}. The one-to-one relationship between peak power consumption and the diversity factor is used to define the second axis on~\figref{fig:diversity_factor}.


With the increase in aggregation level, the maximum load per \ac{CP} reduces for all the cases. When aggregation levels are small, there is a rapid decrease in maximum power per \ac{CP} and a corresponding increase in diversity. This trend saturates for larger aggregations.

Among the 8 scenarios considered, the highest power levels are observed for dynamic energy prices (DE), and the lowest for the flat energy prices with segmented tariffs (FE-$p^{+/-}$). In all cases, the addition of a segmented network tariff effectively reduces the peak power consumption, with lower power thresholds ($p^-$) and higher fees ($\lambda^+$) being more effective.


\par



\section{Conclusions} \label{Conclusions}
The study analyzes the aggregate peak \ac{EV} charging demand on distribution networks when cost-optimizing \ac{EV} users are exposed to various combinations of network tariffs and energy prices. The results indicate that, when users are not exposed to varying energy prices, multi-level segmented network tariffs that incentivize flattening of individual demand patterns are also able to effectively flatten aggregate load profiles and reduce demand peaks, as evidenced by the diversity factor and distribution of maximum charging power per \ac{CS} results. In a more realistic setting where the current \ac{EV} users are exposed to dynamic energy prices, our simulation results indicate that the combination of segmented network tariffs and dynamic energy prices is effective in limiting power demand in low-price hours compared to the scenario only with dynamic energy prices. In addition, lower power levels and higher price levels for the middle capacity band (i.e., more restrictive tariffs) are the most favourable configurations for the segmented network tariff. 

In conclusion, peak-limiting tariffs are a promising approach for reducing the challenge posed by peak \ac{EV} charging demand, whether or not they are used in combination with dynamic energy prices. This emphasizes the importance of tariff design in promoting efficient utilization of distribution network resources. Future work will further investigate the sensitivity of results to different data sets and tariff/price parameters. 
\balance

\bibliographystyle{IEEEtran}
\bibliography{ref}
\vspace{12pt}
\color{red}

\end{document}